\documentclass[pra,twocolumn,superscriptaddress,notitlepage,nofootinbib]{revtex4-2}
\DeclareMathAlphabet\mathbfcal{OMS}{cmsy}{b}{n}
\usepackage[utf8]{inputenc}
\usepackage[T1]{fontenc}
\usepackage[english]{babel}
\usepackage[dvipsnames]{xcolor}
\usepackage{lmodern,fancyhdr}
\usepackage{graphicx,float}
\usepackage{enumitem,amsmath,amssymb,amsthm,bm,array,mathdots,mathrsfs,dsfont,bbold}
\usepackage[colorlinks]{hyperref}
\hypersetup{colorlinks,linkcolor=blue,citecolor=blue,urlcolor=blue,final}
	
\usepackage{lineno}

 \usepackage{url}
 \usepackage{siunitx}
 \usepackage{braket}
 \usepackage[normalem]{ulem}
 \usepackage{placeins}

\newcommand{\expu}{{\rm e}}

\newcommand{\polangle}{\theta_p}
\newcommand{\myeqref}{Eq.~\eqref}
\newcommand{\myeqsref}{Eqs.~\eqref}
\newcommand{\asym}{{\zeta}}
\newcommand{\camangle}{\theta_c}
\newcommand{\HWPangle}{\theta_H}
\newcommand{\initanlge}{\theta_0}

\usepackage{scalefnt} 

\newcommand{\affil}{Department of Physics, Humboldt-Universität zu Berlin, 10099 Berlin, Germany}

\begin{document}
\scalefont{1.05}
\title{Measuring deviations from a perfectly circular cross-section \\of an optical nanofiber at the Ångström scale}

\author{Jihao Jia}
\affiliation{\affil}
\author{Felix Tebbenjohanns}
\affiliation{\affil}
\author{Thomas Hoinkes}
\affiliation{\affil}
\author{J{\"u}rgen Volz}
\affiliation{\affil}
\author{Arno Rauschenbeutel}
\affiliation{\affil}
\author{Philipp Schneeweiss}
\email{philipp.schneeweiss@hu-berlin.de}
\affiliation{\affil}

\begin{abstract}
Tapered optical fibers (TOFs) with sub-wavelength-diameter waists, known as optical nanofibers, are powerful tools for interfacing quantum emitters and nanophotonics. 
These applications demand stable polarization of the fiber–guided light field. 
However, the linear birefringence resulting from Ångström-scale deviations in the nanofiber’s ideally circular cross-section can lead to significant polarization changes within millimeters of light propagation.
Here, we experimentally investigate such deviations using two in-situ approaches. 
First, we measure the resonance frequencies of hundreds of flexural modes along the nanofiber, which exhibit splitting due to the non-circular cross section.
By analyzing the mean resonance frequencies of each pair and the corresponding frequency splitting, we conclude that the nanofiber can be well described as having an elliptical cross-section with a mean radius of \SI{255.6(9)}{nm}, where the semi-axes differ by only about $\SI{2}{\angstrom}$.
Second, we monitor the polarization of the guided light field by imaging the light scattered out of the nanofiber and observe a periodic polarization change along it.
From the linear birefringence due to the elliptical cross-section, we infer a comparable difference in the semi-axes as the first method, and determine the orientation of the polarization eigenaxes.
Our work is crucial for any fundamental or applied study that requires a well-controlled interaction between guided light and matter, in particular for quantum memories, frequency conversion, or lasing that require a large interaction length.
\end{abstract}

\date\today

\maketitle

\paragraph{Introduction.}

Optical nanofibers can be realized as the waist of tapered optical fibers (TOFs), and are powerful experimental tools~\cite{Tong2012}, e.g., to interface guided light with emitters such as NV centers, cryogenically cooled molecules, or laser-cooled atoms~\cite{Morrissey2013, Solano2017, Nayak2018}. 
The strong transverse confinement of guided light in nanofibers leads to high intensities~\cite{Le2004}, rendering this system well-suited also for non-linear optics experiments~\cite{Birks2000, Spillane2008, Wiedemann2010, Zhang2023}.
Other areas where TOFs find applications are in fiber-based beam splitters~\cite{Keiser2021}, active components such as fiber null-couplers~\cite{Birks1996, Blaha2024}, and particle trapping~\cite{Praveen2023}.

The manufacture of TOFs has matured over the last decades, enabling, e.g., power transmissions in excess of 99~\%~\cite{Nagai2014, Hoffman2014}, controlled excitations of higher-order modes guided in the nanofiber part~\cite{Frawley2012, Ravets2013}, a high accuracy of the radius profile of the produced TOF~\cite{Warken2004, Sumetsky2006, Wiedemann2010, Madsen2016, Fatemi2017}, and excellent mechanical properties, e.g., sustaining GPa-scale tensile strains~\cite{Holleis2014}. 
In most works thus far, the cross-section of TOFs is assumed to be circular, which appears plausible as a result of the surface tension aiming to minimize the nanofiber's circumference during the heat-and-pull production~\cite{Birks1992, Ward2014}. 
Remarkably, if the nanofiber had a slightly elliptical cross section with one semi-axis being only a few Ångström larger than the other, the polarization of the guided light exhibits a periodic modulation after only a few millimeters of propagation~\cite{Goban2012}, due to the resulting linear birefringence of the nanofiber wave\-guide. 
A characterization of TOFs and their cross-section with Ångström-scale resolution has, however, not been available until very recently~\cite{Fatemi2017, Wang2025}.

Here, we experimentally determine Ångström-scale deviations from the ideally circular cross-section of optical nanofibers using two methods.
The first one relies on measuring mechanical resonances of the flexural modes, which appear as frequency-split pairs.
From the average frequencies of each mode pair, we extract the mean radius of the nanofiber, while the splitting reveals the difference between the ellipse’s semi-axes.
Moreover, we introduce a general asymmetry parameter for TOFs with elliptical cross-sections and analytically link it to the frequencies of the two split mechanical modes, valid also if the mean radius of the TOF changes along the fiber. 
This well describes our experimental observations across hundreds of modes. 
Our second method relies on the characterization of light scattered out of the nanofiber, which reveals the local polarization of the guided light. 
We provide an analytic formula relating the nanofiber's asymmetry parameter and its linear birefringence, yielding comparable deviations from the ideally circular nanofiber cross-section as the first method. 
Moreover, the optical method allows determining the orientation of the two polarization eigen axes of the linearly birefringent nanofiber with respect to the lab frame. 

\begin{figure}[t] 
\centering
{\includegraphics[width=.9\linewidth]{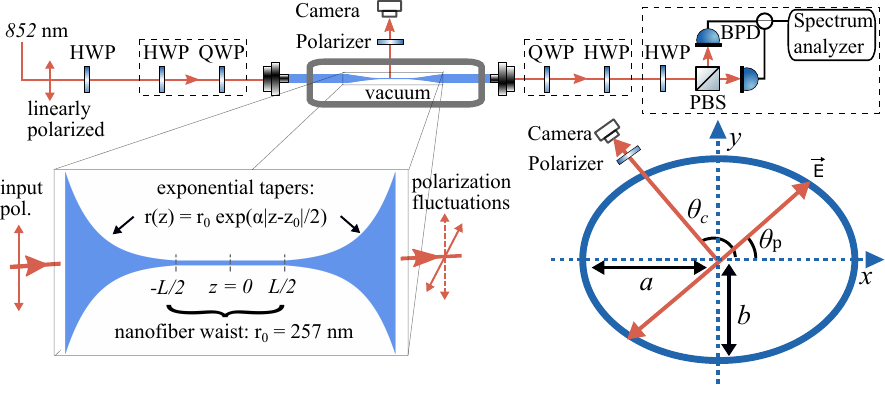}}
\caption{Schematic of the experimental setup. 
We launch linearly polarized laser light into the TOF, which is placed inside a vacuum chamber.
The radius profile of the TOF, including the exponential tapers and the nanofiber waist, is illustrated in the bottom left.
The transmitted light is directed into a balanced detection setup to measure polarization fluctuations induced by the mechanical motion of the TOF.
We monitor the scattering from the nanofiber waist using a camera outside the chamber while controlling the polarization along the nanofiber by the wave plates in front.
In the bottom right, we sketch the elliptical cross-section of the nanofiber with two semi-axes, $a$ and $b$. 
The camera is aligned at an angle $\camangle$ with respect to the semi-major axis, while the polarization angle is $\polangle$.
BPD: Balanced photodiode, HWP: half-wave plate, QWP: quarter-wave plate,  PBS: polarizing beamsplitter.
}
\label{fig:setup}
\end{figure}

\paragraph{Experimental setup.}
Our experimental setup is depicted in Fig.~\ref{fig:setup}. 
We mount a TOF with a nanofiber waist in a vacuum chamber with a typical pressure of \SI{2e-8}{mbar}.
As illustrated in the inset of Fig.~\ref{fig:setup}, the nanofiber waist has a nominal radius of $r_0 = \SI{257}{nm}$
\clearpage
\onecolumngrid 

\twocolumngrid 
\clearpage
\noindent
over a length of $L = \SI{10}{mm}$.
The two ends of the nanofiber waist are connected to exponential tapers that extend from the waist radius $r_0$
up to approximately \SI{9}{\micro m}, following the profile $r(z) = r_0 \exp(\alpha |z-z_0|/2)$, with $\alpha/2 \approx \SI{0.38}{mm^{-1}}$ and $z_0 = \mp L/2$ for the left and right taper, respectively.
We launch linearly polarized laser light with a wavelength of $\lambda = \SI{852}{nm}$ into the TOF.
The TOF is manufactured from a commercial single-mode fiber (SM800-5.6-125) in a heat-and-pull process~\cite{Birks1992, Ward2014}.
The birefringence between the input/output fiber ends and the nanofiber waist is prior unknown. 
To compensate, we place a half-wave plate (HWP) and a quarter-wave plate (QWP) before and after the fiber.
A camera outside the vacuum chamber images the light scattered from the nanofiber waist.
We define the fiber axis as the 
$z$‑axis and place a polarizer before the camera, aligned to block the longitudinal component of the electrical field $E_z$~\cite{Vetsch2012}.
More details will be discussed later.
In transmission, a polarization analysis setup, consisting of a HWP, a polarizing beamsplitter (PBS), and a balanced photodiode (BPD), measures the polarization fluctuations.

\begin{figure}[htb] 
\centering
{\includegraphics[width=\linewidth]{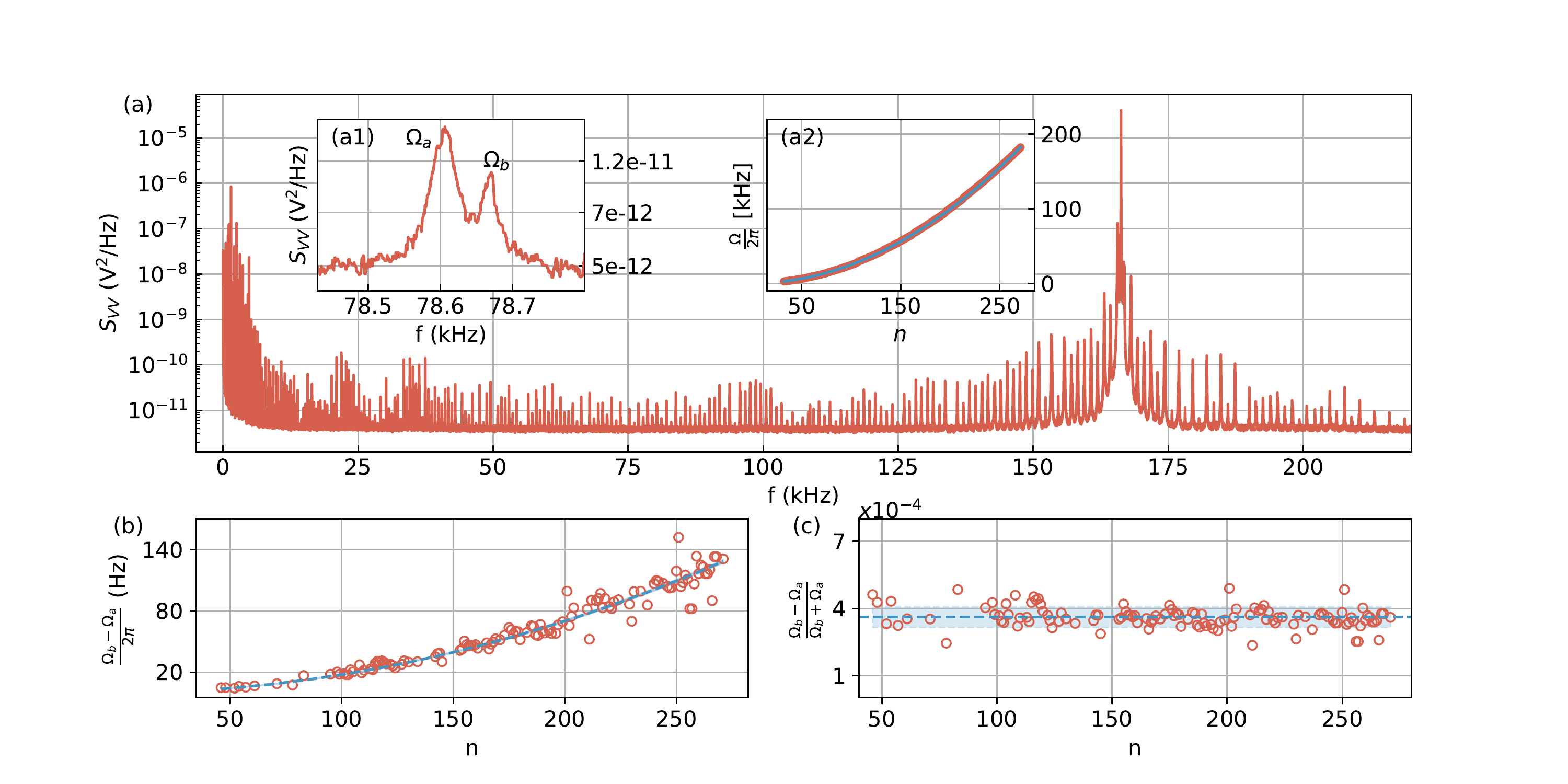}}
\caption{Measurement and analysis of flexural modes, focusing on mode splitting.
(a) Power spectral density (PSD), $S_{VV}(f)$, at thermal equilibrium.
The fundamental torsional mode at \SI{166.3}{kHz} appears as the most prominent peak, while hundreds of flexural modes manifest themselves as smaller peaks in the PSD.
Inset (a1): a zoomed-in view of the PSD reveals a clear splitting of a single flexural mode.
Inset (a2): the average resonance frequency, $\Omega = \frac{\Omega_a + \Omega_b}{2}$, as a function of the mode number $n$.
To these data points (red dots), we fit with \myeqref{eq:resonance_r} (blue line) to determine the mode number of the first captured peak, and the effective length $L_\text{eff}$.
(b) Mode splitting, $\frac{\Omega_b - \Omega_a}{2 \pi}$, as a function of the mode number $n$.
Red dots are the data points while the blue line is a fit according to \myeqref{eq:splitting}, yielding $a-b =\SI{0.180(2)}{nm}$.
(c) The normalized mode splitting, $\frac{\Omega_b - \Omega_a}{\Omega_a + \Omega_b}$, as a function of the mode number, $n$, remains constant.
The blue dashed line shows the average value of the data points (red dots), corresponding to an asymmetry factor of $\SI{3.62(5)e{-4}}{}$.
The light blue area represents the standard deviation of the data points.
}
\label{fig:PSD}
\end{figure}

\paragraph{Flexural modes.}
In Fig.~\ref{fig:PSD}(a), we present the power spectral density (PSD), $S_{VV}(f)$, of the signal measured with the balanced photodiode in the polarization analysis setup recorded with a spectrum analyzer, while the system is in equilibrium with its environment.
The PSD covers a frequency range from \SI{0}{kHz} to \SI{220}{kHz} with a frequency resolution of \SI{0.4}{Hz}.
The dominant peak at \SI{166.3}{kHz} corresponds to the fundamental torsional mode of the nanofiber~\cite{Wuttke13Optically, Tebbenjohanns2023, Su2023}. 
Upon fabrication, the TOF is glued across two separated fiber-holder halves on a translation stage and initially stretched, yielding only torsional modes in the PSD.
While monitoring the frequency shift of the torsional mode, we gradually relax the TOF to reduce the strain until it is close to zero, at which point the flexural modes appear in the spectrum as numerous peaks~\cite{Song2024}.
As shown in the inset (a1) of Fig.~\ref{fig:PSD}(a), each flexural mode exhibits two resonant peaks.
A simple extension beyond a circular cross-section is to treat our nanofiber as a cylinder with a slightly elliptical cross-section, defined by semi-major axis $a$ and semi-minor axis $b$.
Since the flexural modes penetrate into the tapered regions, we model the entire structure as a beam with slightly elliptical cross-section, connected to two exponential tapers.
Remarkably, the resulting resonance frequencies follow the familiar formula for a simple beam with fixed-fixed boundary conditions~\cite{Rao2017} 
\begin{subequations}    \label{eq:resonance}    \begin{align}
    \Omega_{a, b}= & \left( n + \frac{1}{2} \right)^2 \frac{\pi^2}{L_\text{eff}^2} \sqrt{\frac{E I_{a, b}}{\rho A}} \\
    \stackrel{\text{circular}}{=}& \left( n + \frac{1}{2} \right)^2 \frac{\pi^2}{2 L_\text{eff}^2}\sqrt{\frac{E}{\rho}} \, r_0 \label{eq:resonance_r}. 
\end{align}
\end{subequations}
\clearpage

\onecolumngrid 

\twocolumngrid 
\noindent In these expressions, we absorb all effects of the taper’s varying radius into a single effective length,
\begin{equation}
    L_\text{eff} \approx L + 8 / \alpha,
\end{equation}
which in our case yields $ L_\text{eff} \approx \SI{20.52}{mm}$.
In \myeqref{eq:resonance}, $\Omega_{a, b}$ are the angular resonance frequencies for the oscillations bending about the respective semi-axis, $n$ is the mode number, $E$ is Young's modulus, $\rho$ is the mass density, $A$ is the cross-sectional area, and $I_{a,b}$ are the second moments of area, namely $I_a = A \, b^2/4$ and $I_b = A \, a^2/4$.
Flexural modes bending about the semi-major axis have a different second moment of area than those about the semi-minor axis, leading to the frequency splitting for each mode pair
\begin{equation}
    \Omega_{b} - \Omega_{a} = \left( n + \frac{1}{2} \right)^2 \frac{\pi^2}{2L_\text{eff}^2} \sqrt{\frac{E}{\rho}}(a-b).\label{eq:splitting}
\end{equation}
Based on the data shown in Fig.~\ref{fig:PSD}(a), we identify the resonance frequencies from approximately \SI{3}{kHz} onward and record the relative mode number, $n_\text{rel}$, with respect to the first observed peak.
We take the mean value of each double peak and fit the resonance frequencies using \myeqref{eq:resonance_r} (inset (a2) of Fig.~\ref{fig:PSD}(a)), assuming a circular cross-section with mean radius $r_0$ and $I_a = I_b = \frac{\pi}{4} r_0^4$.
From the fit, we determine the absolute mode number as $n = n_\text{rel} + n_0$, where $n_0 = \SI{27.3(2)}{} \approx 27$ is the fit parameter corresponding to the mode number of the first observed peak.
We also extract the mean radius of the nanofiber as $r_0 = \SI{255.6(9)}{nm}$, closely matching the nominal value of \SI{257}{nm}.
Second, we calculate the mode splitting, $\Omega_b - \Omega_a$, and fit it using \myeqref{eq:splitting}, as shown in Fig.~\ref{fig:PSD}(b).
The fitting yields $a-b \approx \SI{0.180(2)}{nm}$, 
which is much smaller than the mean radius. 
To confirm this small deviation, we introduce the ``asymmetry'' parameter $\asym = \frac{a-b}{a+b} \ll 1$, assume it remains constant even in the tapers, and relate it to the frequency splitting according to (see Appendix~\ref{App:A})
\begin{align} \label{eq:asym}
    \asym = \frac{a-b}{a+b} \approx \frac{\Omega_b - \Omega_a}{\Omega_a + \Omega_b}.
\end{align}
Remarkably, this expression is independent of other fit parameters (e.g., $L_\text{eff}$ and $n$) as well as material properties (e.g., $\rho$ and $E$).
In Fig.~\ref{fig:PSD}(c), we plot $\frac{\Omega_b - \Omega_a}{\Omega_a + \Omega_b}$ as a function of the mode number $n$.
By taking the mean value of the data points, we obtain $\asym = \SI{3.62(5)e{-4}}{}$, which is in good agreement with the asymmetry parameter calculated from the previously estimated values of $a-b$ and $r_0$.

The asymmetry of the fiber's cross-section not only causes mode splitting in the mechanical modes but also gives rise to linear birefringence with two principal axes for the guided optical mode.
The effective refractive index is larger when the guided light is polarized along the semi-major axis, whereas light polarized along the semi‑minor axis experiences a smaller effective index.
Thereby, the nanofiber section features linear birefringence, characterized by a difference in the propagation constants of the two light fields as (see App.~\ref{App:B}) 
\begin{equation}
    \Delta \beta = \beta_b - \beta_a = \frac{2 \pi}{\lambda} \times 0.176 \ \asym = \SI{0.442(9)}{mm^{-1}}.
\end{equation}
Here, the factor $0.176$ is specific to the parameters, $\lambda =\SI{852}{nm}$, $r_0 = \SI{257}{nm}$, and a refractive index of the fiber material of $n = 1.452$, while $\beta_a$ and $\beta_b$ denote the propagation constants for light polarized along the eigenaxes $a$ and $b$, respectively.
In this case, an initially linearly polarized input light field whose polarization is not aligned to one of the polarization eigenaxes undergoes periodic changes in its polarization state.
During the propagation through the fiber, the light also assumes elliptical and circular polarization states.
After a so-called beat length of $L_\text{beat}= 2 \pi / \Delta \beta = \SI{14.2(3)}{mm}$, the light field goes back to the initial linear polarization state.

\begin{figure}[t]
\centering
{\includegraphics[width=\linewidth]{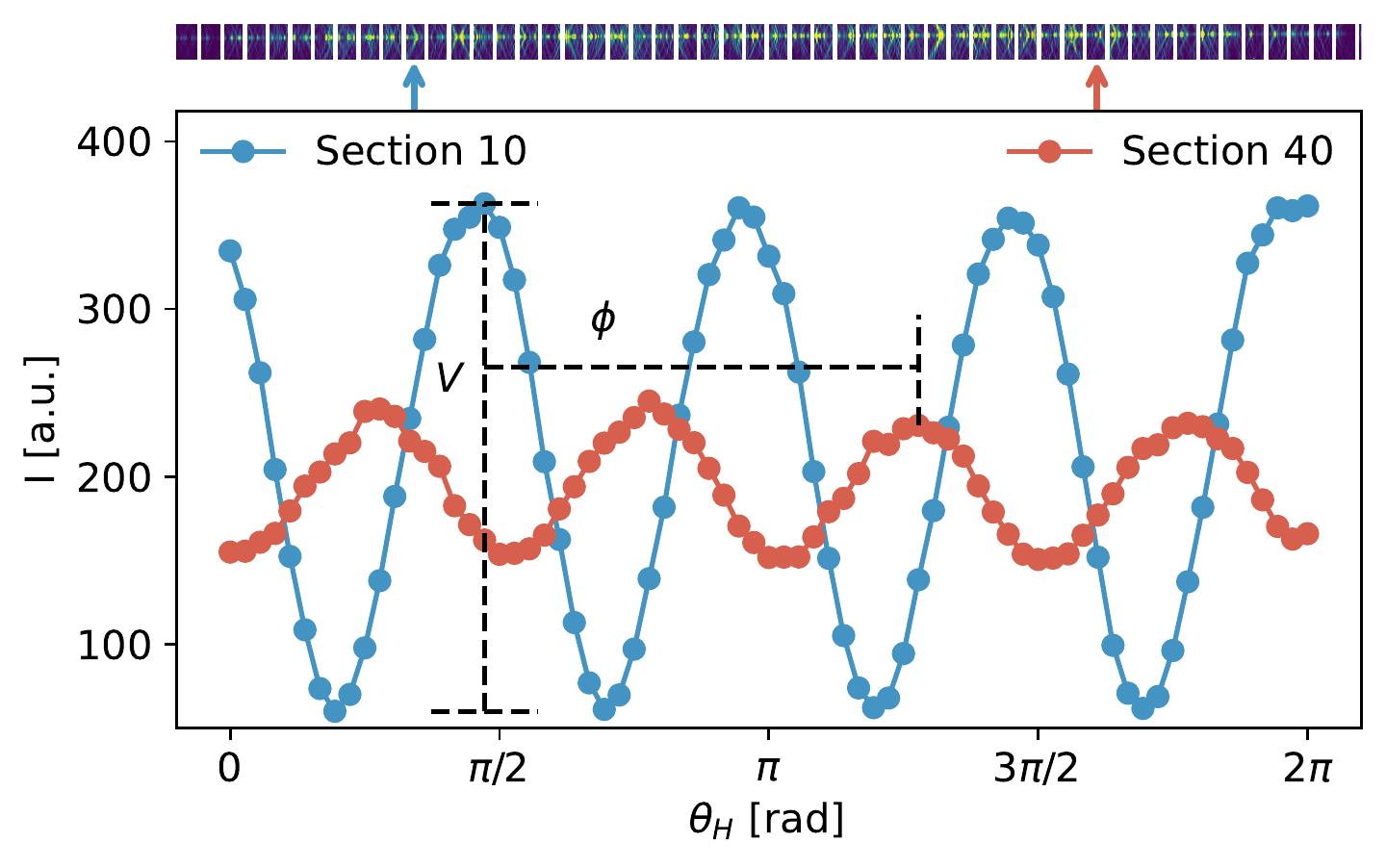}}
\caption{
Polarization dependent scattering pattern of the nanofiber.
The top panel shows an image of the nanofiber averaged over $\theta_H$, see main text. 
The horizontal bright line corresponds to the scattering from the nanofiber while the white vertical lines divide the nanofiber into multiple sections.
The bottom panel presents the intensities of sections 10 and 40, showing periodic modulation as a function of $\HWPangle$.
From this modulation, we extract the visibility and phase of the scattered light.
}
\label{fig:intensity}
\end{figure}

\begin{figure}[ht]
\centering
{\includegraphics[width=\linewidth]{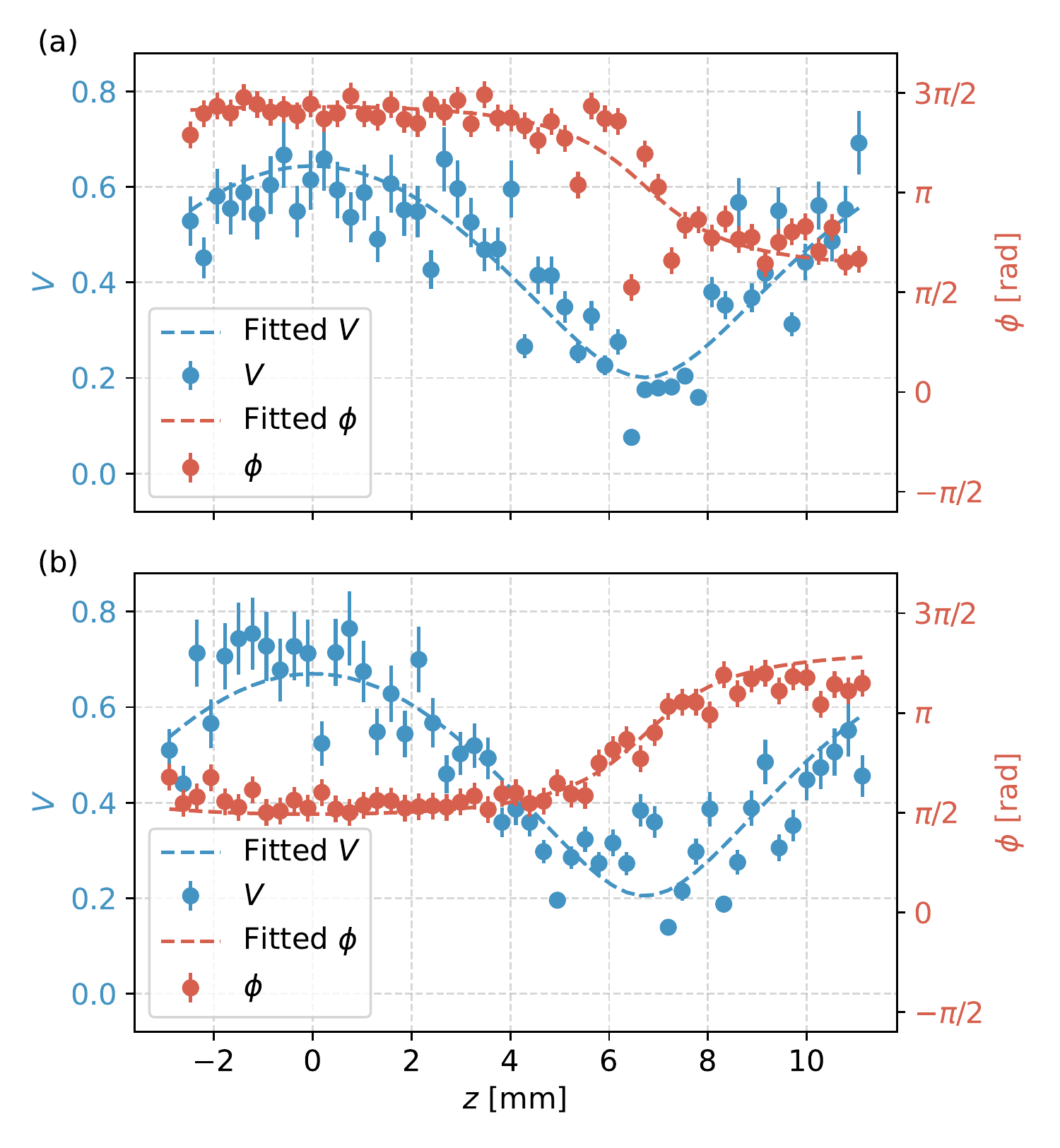}}
\caption{Visibility and phase measurement for camera 1 (a), and camera 2 (b). 
From a joint fit of the visibility data for both cameras using \myeqref{eq:V_phi}, we extract the linear birefringence, $\Delta \beta = \SI{0.222(9)}{mm^{-1}}$, and camera angles, $\theta_{c,1} = \SI{129.9(4)}{^\circ}$ and $\theta_{c,2} = \SI{53.9(9)}{^\circ}$.
The phase is then fitted using the obtained value for $\Delta \beta$.
}
\label{fig:visibility}
\end{figure}

\paragraph{Polarization imaging.}
The idea of the polarization imaging is based on Refs.~\cite{Vetsch2012, Joos2019}. 
We analyze the light scattered out of the nanofiber by surface roughness and impurities, and density inhomogeneities in the bulk silica. 
Assuming Rayleigh scatterers, they are expected to preserve the local polarization of the guided light. 
Following the dipole radiation pattern, maximum intensity is expected when the optical axis of the imaging system is perpendicular to the local linear polarization of the guided light, while minimum intensity is expected when they are parallel.
The resulting image recorded with the camera is shown in Fig.~\ref{fig:intensity} (top), where the horizontal bright line corresponds to the light scattered by the nanofiber. 
We then divide this region of interest into 51 sections, as indicated by the white vertical lines, and record the integrated intensity $I$.
We adjust the first HWP at the input to rotate the polarization of the guided field and measure the visibility of the integrated intensity, which is defined as $V = \frac{I_\text{max} - I_\text{min}}{I_\text{max} + I_\text{min}}$.
We then tune a second HWP and a QWP before the fiber to maximize the visibility at a specific section, where we set $z=0$ and assume the light field is linearly polarized~\cite{Vetsch2012, Joos2019}.
At that position the polarization angle follows $\polangle = \initanlge + 2 \HWPangle$~\cite{Joos2019}, where $\initanlge$ is the initial polarization angle and $\HWPangle$ is the rotation angle of the first HWP.
In Fig.~\ref{fig:intensity} (bottom), we display the intensities, $I$, of sections 10 and 40 as a function of $\HWPangle$.
The intensity in each section exhibits a periodic modulation in response to the HWP rotation, allowing us to extract the visibility and relative phase (with respect to $\HWPangle$) at every section.
In App.~\ref{App:C}, we attribute this behavior to the linear birefringence of the nanofiber as a results of an elliptical cross-section, where the principal axes for the guided optical modes align with the two semi-axes.
As sketched in Fig.~\ref{fig:setup} (bottom right), we define the ``fiber frame'', such that the x-axis is oriented along the semi-major axis and the y-axis along the semi-minor axis.
The predicted intensity is
\begin{align}\label{eq:intensity}
    I (\camangle, \polangle, z) \propto 1 - \text{Re}\{e^{i 2 \polangle} W\},
\end{align}
where $\camangle$ is the camera angle, $z$ is the position along the fiber axis, and $W= \cos 2 \camangle - i \sin 2 \camangle \cos \Delta \beta z$.
The visibility and phase are then given as
\begin{align} \label{eq:V_phi}
    V_W = |W|,\, \phi_W = \text{Arg} (W),
\end{align}
where \text{Arg} denotes the argument of a complex number.
At each position $z$, the visibility and phase depend only on two factors: the camera angle $\camangle$ and the linear birefringence $\Delta \beta$.
Initially, prior to the measurement, $\camangle$ is unknown.
If the camera happens to be aligned along one of the two principal axes such that $\camangle=0$, the measured signal will become independent of $\Delta \beta$.
To avoid this scenario, we mount two cameras at different angles to do the intensity measurement, with an angle separation of $\Delta \camangle = \SI{72}{^\circ} \neq \SI{90}{^\circ}$.
In Fig.~\ref{fig:visibility}, we plot the visibility and phase recorded by camera 1 and camera 2, respectively.
By simultaneously fitting the visibility data from both cameras using \myeqref{eq:V_phi}, we find that $\Delta \beta = \SI{0.222(9)}{mm^{-1}}$, $L_\text{beat} = \SI{28.2(12)}{mm}$, $a-b \approx \SI{ 0.085(4)}{nm}$, and the camera angles, $\theta_{c,1} = \SI{129.9(4)}{^\circ}$ and $\theta_{c,2} = \SI{57.9(4)}{^\circ}$.
We infer that the angle between the semi-major axis and the horizontal plane is about \SI{36}{^\circ}.
The phase data is then fitted while fixing $\Delta \beta$ to the above value. 

We briefly compare the findings obtained with the flexural mode and the polarization imaging method. 
In both cases, we find Ångström-scale deviations
from the commonly assumed circular cross-section of the nanofiber. 
All experimental data can be well described assuming a TOF with an elliptical cross-section for which the relative deviation of the semi-major and the semi-minor axis is at the $10^{-4}$ level. 
Although it is remarkable that such small deviations can be detected, the actual values found with the two methods differ by about a factor of 2.

We speculate that this deviation arises from adsorption and desorption processes, which, depending on the conditions (pressure in the vacuum chamber, power and polarization of guided light, etc.) experienced by the nanofiber prior to the measurement, may add and remove adsorbates on the nanofiber surface. 
This hypothesis is compatible with the observation that the mode splitting observed in the flexural mode measurement always stays approximately constant and agrees within better than 10~\% when measured for different pressures in the vacuum chamber or different probe laser powers: 
It is easy to show that the asymmetry parameter $\zeta$ does not change when adsorbates build up on the nanofiber surface, i.e., when material is deposited that adds to the nanofibers inertia but can be assumed not to alter the restoring force that underlies the flexural mode vibrations. 
The beat length found in the polarization imaging and, thus, the inferred asymmetry of the nanofiber, however, scatters by about 10~\% when carried out at the same pressure and by about 50~\% when markedly varying pressures or probe laser powers (and thus the equilibium temperature of the nanofiber).
Surface coverage of the nanofiber may change under these different conditions, thereby modifying its birefringence in a way that is compatible with the observed variations in beat length~\cite{Delic2019, Ricci2022}.

\paragraph{Conclusions.}
In conclusion, we demonstrated two experimental methods that allow us to characterize the elliptical cross-section of the nanofiber waist of a TOF, with immediate relevance for defining and maintaining a well-defined polarization of the guided light in optical nanofibers. 
Both methods, one based on analyzing the frequency splitting of flexural mode mechanical resonances and the other based on analyzing the polarization of the light scattered out of the fiber, yield Ångström-scale deviations from an ideally circular fiber cross-section. 
Moreover, both methods are simple to implement, work with any tapered optical fiber, and are compatible with in-vacuum applications.

Similar deviations from a perfectly round cross-section are observed for most tapered fibers produced with our fiber pulling rig, and similarly seem to be present for tapered fibers made by other research groups~\cite{Wang2025}. 
In future work, the origin of those imperfections will be investigated with the aim of controlling the exact shape during the fabrication. 
For example, more strongly elliptical fiber cross-sections appear as a promising approach in order to create polarization-maintaining optical nanofibers~\cite{Jung2010, Xuan2010}. 
Our work will enable an improvement in tapered-fiber coupling, and in light-matter interaction, e.g., via Brillouin scattering~\cite{Beugnot2014, Florez2016}.
It will also allow a higher level of control in interfacing, e.g., plasmonic particles~\cite{Petersen2014} and quantum emitters such as laser-cooled atoms~\cite{Vetsch2012, Goban2012, Lee2015, Ruddell2017, Ostfeldt2017, Corzo2019, Gupta2022, Kestler2023}.
From another perspective, shape control during fabrication may also allow optimization of the optomechanical coupling strength for both flexural and torsional modes~\cite{Tebbenjohanns2023, Su2023, Pennetta2020, Aspelmeyer2014}. 
This could offer the possibility to cool the mechanical motions to the quantum regime, making TOFs a promising platform for hybrid quantum systems~\cite{Treutlein2014}.

\begin{acknowledgments}
We thank Riccardo Pennetta for conducting numerical simulations;
Wolfgang Alt, Johannes Piotrowski and Leonid Yatsenko for their insightful discussions.
\end{acknowledgments}

\appendix
\section{Lateral vibrations of slightly elliptical beams} \label{App:A}
The equation of motion for the free lateral vibration of a homogeneous beam oriented along the z-axis with non-uniform cross-section is~\cite{Rao2017}
\begin{equation} \label{eq:DEQ_Rao}
   \partial_z^2[EI(z)\partial_z^2 w(z,t)] + \rho A(z)\partial_t^2 w(z,t) = 0,
\end{equation}
where $E$ is Young's modulus, $I(z)$ is the second moment of area, $w(z,t)$ is the deflection, $\rho$ is the mass density, and $A(z)$ is the area of the cross-section.
We assume that the beam has an elliptical cross-section with semi-axes $a(z)$ and $b(z)$, and define the asymmetry as $\asym= \frac{a-b}{a+b}$, such that
\begin{subequations}
    \begin{align}
        a(z) &= r_0 r(z)[1 + \asym(z)], \\
        b(z) &= r_0 r(z)[1 - \asym(z)],
    \end{align}
\end{subequations}
where $r_0$ is the characteristic radius of the structure and $r(z)$ is the (unit-less) radius profile.
For $\asym \ll 1$, we find
\begin{subequations}\label{eq:A_I}
    \begin{align}
        A(z) &= \pi a(z)b(z)=\pi r_0^2 r(z)^2 + O(\asym^2)\notag\\ 
        &= A_0 r(z)^2 + O(\asym^2),  \\
        I_a(z) &= \frac{A(z)}{4} b(z)^2 = \frac{\pi r_0^4}{4} r(z)^4 [1 - 2\asym(z)] + O(\asym^2) \notag\\
        &= I_0 r(z)^4 [1 - 2\asym(z)] + O(\asym^2),\\
        I_b(z) &= \frac{A(z)}{4} a(z)^2 = \frac{\pi r_0^4}{4} r(z)^4 [1 + 2\asym(z)] + O(\asym^2)\notag \\
        &= I_0 r(z)^4 [1 + 2\asym(z)] + O(\asym^2),
    \end{align}
\end{subequations}
where $A_0 = \pi r_0^2$ and $I_0 = \frac{\pi r_0^4}{4}$.
Substituting \myeqsref{eq:A_I} into \myeqref{eq:DEQ_Rao}, we obtain
\begin{equation}
    \frac{c_0^2}{r(z)^2}\partial_z^2 [r(z)^4[1 \mp 2\asym(z)]\partial_z^2 w(z,t)] = -\partial_t^2 w(z,t),
\end{equation}
where $c_0^2 = \frac{E I_0}{\rho A_0}$. The solution can be found using the method of separation of variables,
\begin{equation}
    w(z,t) = W(z)T(t).
\end{equation}
This leads to 
\begin{align} \label{eq:deq_sep_var}
    &\frac{1}{W(z)} \frac{c_0^2}{r(z)^2} \, \partial_z^2 \Big(
        r(z)^4 [1 \mp 2\asym(z)] \, \partial_z^2 W(z) 
    \Big) \notag \\
    &= -\frac{\partial_t^2 T(t)}{T(t)} = \Omega^2,
\end{align}
where any solution $\Omega$ is an angular resonance frequency.
\myeqref{eq:deq_sep_var} can be written as two equations,
\begin{subequations}
    \begin{align}
        & \partial_t^2T(t)+\Omega^2T(t)=0, \\
        & \frac{1}{r(z)^2} \partial_z^2 [r(z)^4[1\mp2\asym(z)] \partial_z^2W(z)] - \frac{\Omega^2}{c_0^2}W(z)=0.
    \end{align}
\end{subequations}
For constant asymmetry, $\asym(z) = \asym$, we will find
\begin{subequations}
    \begin{align}\label{eq:deq_beta}
    &\frac{1}{r(z)^2}\partial_z^2[r(z)^4\partial_z^2W(z)]-s^4W(z)=0, \\
    &\text{with }
    s = \sqrt[4]{\frac{\Omega^2}{c_0^2(1\mp2\asym)}} \approx \sqrt{\frac{\Omega}{c_0(1\pm\asym)}}.
\end{align}
\end{subequations}
For any beam, there exists an infinite set of normal modes, each associated with a distinct resonance frequency~\cite{Rao2017}.
Accordingly, we assume the solutions to \myeqref{eq:deq_beta} are given by $s = s_n$, where $ n = 1, 2, 3, \ldots$ is the mode number, which yields
\begin{equation}\label{eq:asym_appendix}
        \Omega_n = s_n^2c_0(1\pm\asym) \Rightarrow
        \frac{\Omega_b - \Omega_a}{\Omega_a + \Omega_b} = \asym.
\end{equation}
Despite the inhomogeneous radius profile of the beam, as long as $\asym$ stays constant along the beam, the normalized mode splitting will always follow \myeqref{eq:asym_appendix}.
This is important for our analysis, as the flexural modes propagate not only through the nanofiber waist but also into the exponential tapers, where the radius varies continuously.

\section{Linear birefringence for a dielectric waveguide with a slightly elliptical cross-section} \label{App:B}
The analysis of this section follows Ref.~\cite{Yeh2008}.
We define the elliptical coordinates $(\xi, \eta, z)$ as
\begin{subequations}
\begin{align}
    x &= q \cosh \xi \cos \eta, \\
    y &= q \sinh \xi \sin \eta, \\
    z &= z,
\end{align}
\end{subequations}
where $q$ is half the distance between the foci (i.e., the focal length in elliptical coordinates), $\xi\in[0,\infty)$, and $\eta\in[0,2\pi)$.
In these coordinates, $\xi=\xi_0=$ constant defines an ellipse with the semi-axes 
\begin{subequations}
\begin{align}
    a &= q \cosh \xi_0 = \frac{q}{2}\left(\expu^{\xi_0}+\expu^{-\xi_0}\right), \\
    b &= q \sinh \xi_0 = \frac{q}{2}\left(\expu^{\xi_0}-\expu^{-\xi_0}\right).
\end{align}
\end{subequations}
We then find the asymmetry as
\begin{equation}
    \asym = \frac{a-b}{a+b} = \expu^{-2\xi_0}.
\end{equation}
For $\xi_0 \to \infty$, the shape approaches a circle, but the area,
\begin{equation}
    A = \pi a b = \pi \frac{q^2}{4}\left(\frac{1}{\asym} + \asym\right),
\end{equation}

\noindent diverges in this limit. 
For a finite circle of area \mbox{$A = \pi r_0^2$}, we thus additionally require that
\begin{equation}
    \frac{q^2}{4} = r_0^2 \frac{\asym}{1 + \asym^2} = r_0^2 \asym (1+\mathcal{O}(\asym^2)).
\end{equation}
For an elliptical waveguide, whose elliptical cross-section is parameterized by $q$ and $\xi_0$, with relative permittivity $\epsilon_0$ (outside the ellipse) and $\epsilon_1$ (inside the ellipse), the {\it approximate characteristic equations}~\cite{Yeh2008} for the ordinary and extra-ordinary $HE_{nm}$ modes are
\begin{widetext}
    \begin{subequations}\label{eqs:characteristic_elliptical}
\begin{align} 
    &\left[ \frac{1}{\gamma_1^2} \frac{Ce'_n(\xi_0)}{Ce_n(\xi_0)} + \frac{1}{\gamma_0^2} \frac{Fek'_n(\xi_0)}{Fek_n(\xi_0)} \right]
    \left[ \frac{1}{\gamma_1^2} \frac{Se'_n(\xi_0)}{Se_n(\xi_0)} + \frac{\epsilon_0}{\epsilon_1} \frac{1}{\gamma_0^2} \frac{Gek '_n(\xi_0)}{Gek_n(\xi_0)} \right] - n^2 \frac{\left(\gamma_0^2 + \gamma_1^2\right) \left( \gamma_0^2 + \frac{\epsilon_0}{\epsilon_1} \gamma_1^2 \right)}{\gamma_0^4 \gamma_1^4} = 0,\label{eq:characteristic_elliptical_a}\\  
    &\left[ \frac{1}{\gamma_1^2} \frac{Se'_n(\xi_0)}{Se_n(\xi_0)} + \frac{1}{\gamma_0^2} \frac{Gek '_n(\xi_0)}{Gek_n(\xi_0)} \right]
    \left[ \frac{1}{\gamma_1^2} \frac{Ce'_n(\xi_0)}{Ce_n(\xi_0)} + \frac{\epsilon_0}{\epsilon_1} \frac{1}{\gamma_0^2} \frac{Fek'_n(\xi_0)}{Fek_n(\xi_0)} \right]- n^2 \frac{\left(\gamma_0^2 + \gamma_1^2\right) \left( \gamma_0^2 + \frac{\epsilon_0}{\epsilon_1} \gamma_1^2 \right)}{\gamma_0^4 \gamma_1^4} = 0.\label{eq:characteristic_elliptical_b}
\end{align}
\end{subequations}
\end{widetext}
The functions $Se_n, Ce_n, Fek_n,$ and $Gek_n$ are modified (or radial) Mathieu functions~\cite{Mclachlan1947}. 
The parameters $\gamma_0$ and $\gamma_1$ are defined as
\begin{subequations}
\begin{align}
    \gamma_0^2 &= \frac{q^2}{4} \left(\beta^2 - k_0^2\right), \\
    \gamma_1^2 &= \frac{q^2}{4} \left(k_1^2 - \beta^2\right).
\end{align}
\end{subequations}
Here, $k_0 = (\omega/c) \sqrt{\epsilon_0} $ is the wavenumber in the surrounding medium, $k_1 = (\omega/c) \sqrt{\epsilon_1}$ is the wavenumber inside the bulk material, and $\beta$ is the effective wavenumber of the guided mode. 
Similar to the case of a circular cross-section, the only modes without any cut-off frequency are the $HE_{11}$ modes, where $n=m=1$.

In the following, we Taylor-expand Eqs.~\eqref{eqs:characteristic_elliptical} to first order in $\asym$. 
Since these equations scale with $1/\asym^2$ for vanishing $\asym$, we first multiply both sides with $\asym^2$ to define 
\begin{equation}
    f(\asym; \beta) = \asym^2 \times \left[\text{LHS of Eqs.~\eqref{eqs:characteristic_elliptical}}\right].
\end{equation}
With this, we need to solve $f(\asym; \beta)=0$ by finding the zeros as a function of $\beta$. 
As we are interested in the solution to first order in $\asym$, we use the ansatz
\begin{equation}
    \beta(\asym) = \beta_0 + \asym \Delta \beta +\mathcal{O}(\asym^2),
\end{equation}
which leads to 
\begin{align}
    &f(\asym; \beta(\asym)) = f(\asym; \beta_0 + \asym \Delta \beta +\mathcal{O}(\asym^2))\notag  \\
    &= f(0;\beta_0) + \asym[\partial_\asym f(0; \beta_0) + \Delta \beta \partial_\beta f(0; \beta_0) ] + \mathcal{O}(\asym^2)\notag \\
    &= 0. 
\end{align}
The solution ($\beta_0, \Delta \beta$) is thus found by first finding the solution to $f(0;\beta_0) = 0$. This equation will actually reduce to the characteristic equation for a spherical waveguide. Once $\beta_0$ is known, we find $\Delta \beta$ by 
\begin{equation}
    \Delta \beta = -\frac{\partial_\asym f(0; \beta_0)}{\partial_\beta f(0; \beta_0)}.
\end{equation}
We notice that $\Delta \beta$ of the ordinary mode equals $-\Delta \beta$ of the extra-ordinary mode. 
This follows from the fact that $\partial_\asym f(0; \beta_0)$ changes sign from Eq.~\eqref{eq:characteristic_elliptical_a} to Eq.~\eqref{eq:characteristic_elliptical_b}, while ${\partial_\beta f(0; \beta_0)}$ stays the same.
We find $\beta_0$ by numerically solving $f(0,\beta_0)=0$, which yields a single solution if $r_0/\lambda$ is sufficiently small,  such that the cylinder operates as a single-mode waveguide.
We then numerically compute $\Delta \beta = -\partial_\asym f(0; \beta_0) / \partial_\beta f(0; \beta_0)$.
For the parameters $\{ \lambda, r_0, \epsilon_0, \epsilon_1 \} = \{\SI{852}{nm}, \SI{257}{nm}, 1.00^2, 1.452^2 \} $ (the value of $\epsilon_1$ follows from the Sellmeier's equation), we find 
\begin{subequations}
\begin{align}
    n_\text{eff} &= \frac{\beta_0}{2\pi/\lambda} = 1.143, \\
    \Delta n_\text{eff} &= \frac{\beta_b - \beta_a}{2\pi/\lambda } = \frac{-2\Delta \beta}{2\pi/\lambda } \asym = 0.176 \, \asym. \label{eq:prefactor}
\end{align}
\end{subequations}
\begin{figure}[t]
\centering
{\includegraphics[width=\linewidth]{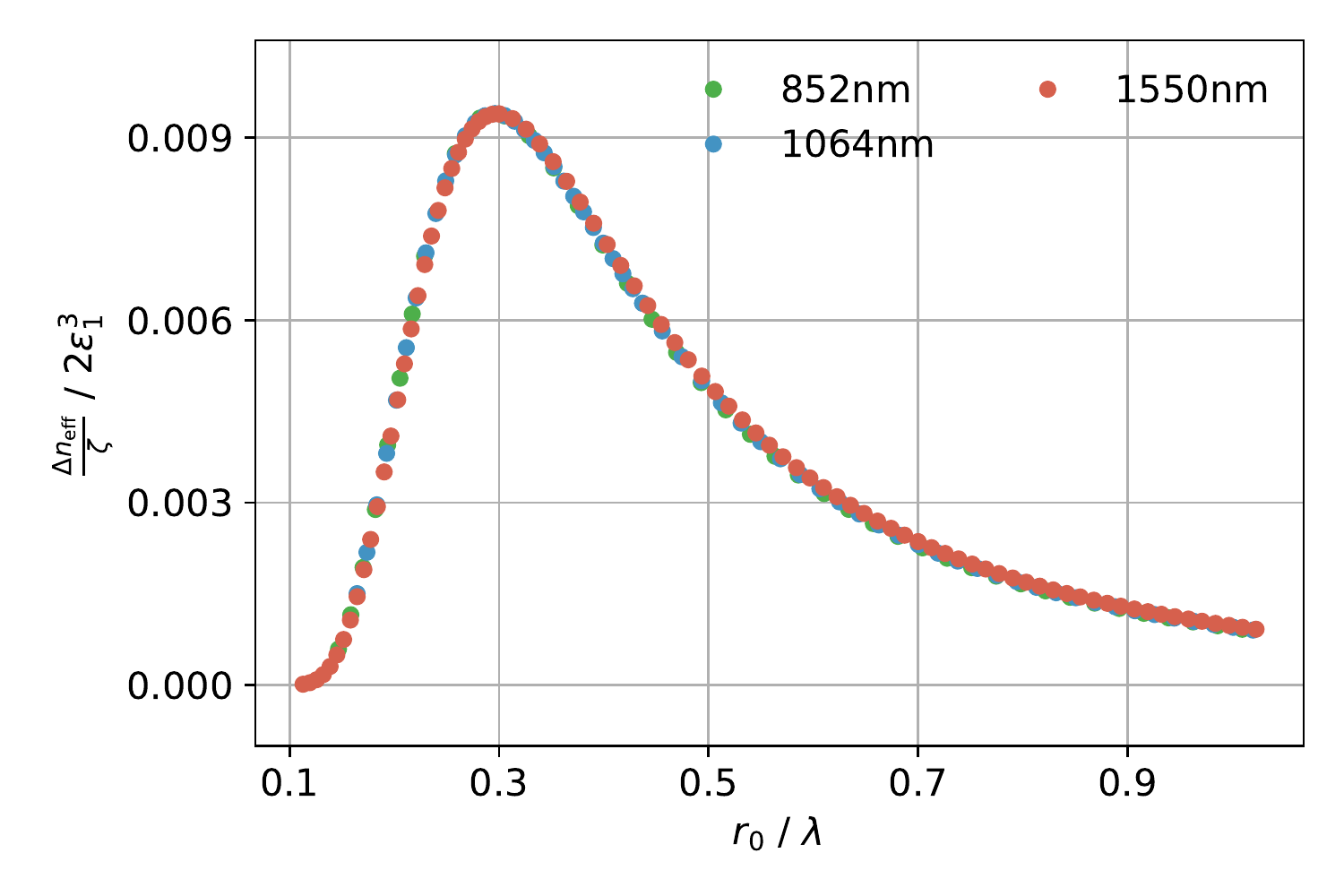}}
\caption{
Differences in the effective refractive indices as a function of $r_0/\lambda$.
The normalized quantity, $\frac{\Delta n_{\text{eff}}}{\asym}/{2\epsilon^3_1}$, allows to relate the geometry of the elliptical cross-section, $\asym$, to the birefringence of the nanofiber in a generally applicable way, i.e., valid for different wavelengths.
Dots in different colors correspond to three wavelengths, \SI{852}{nm}, \SI{1064}{nm}, and \SI{1550}{nm}, respectively.
All calculations are for the fundamental guided mode.
}
\label{fig:prefactor}
\end{figure}
\noindent Figure~\ref{fig:prefactor} shows the normalized prefactor from \myeqref{eq:prefactor}, defined as $\frac{\Delta n_{\text{eff}}}{\asym} / 2 \epsilon_1^3$, plotted against $r_0/\lambda$.
The normalization factor $1 / (2 \epsilon_1^3)$ is an empirical choice that ensures good overlap between the curves at different wavelengths.
Using the data points from Fig.~\ref{fig:prefactor}, one can estimate the birefringence of an optical nanofiber for a given radius, laser wavelength, and asymmetry.

\section{Visibility and phase from the polarization imaging} \label{App:C}
Compared to a linearly polarized light field in the free space, the fundamental $\text{HE}_{11}$ mode in the nanofiber is only quasi-linearly polarized~\cite{Le2004}. 
The guided light has a non-trivial longitudinal component along the fiber axis ($z$-axis), which we filter out by placing a polarizer before the camera.
For the polarization imaging, we approximate the transverse electric field component in the plane orthogonal to the fiber axis as a linear polarization at an angle $\polangle$ at $z=0$, 
as sketched in Fig.~\ref{fig:setup} (bottom right).
This allows us to define the normalized electric field vector as $\Vec{u}(z=0) = \frac{\Vec{E}(z=0)}{|\Vec{E}(z=0)|} = \begin{pmatrix} \cos \polangle \\ \sin \polangle\end{pmatrix}$, which is unitless and has unit amplitude.
Due to the linear birefringence of the nanofiber, the electric field at position $z$ is then
\begin{equation}
    \Vec{u}(z) = \begin{pmatrix} e^{i \beta_a z} \cos \polangle \\ e^{i \beta_b z}  \sin \polangle\end{pmatrix} 
    = e^{i \beta_a z} \begin{pmatrix} \cos \polangle \\ e^{i \Delta \beta z}  \sin \polangle\end{pmatrix}.
\end{equation}
The camera is at an angle of $\camangle$, such that the camera axis is $\begin{pmatrix}  \cos \camangle \\   \sin \camangle\end{pmatrix}$.
The projection of the electrical field onto the direction orthogonal to the camera axis is
\begin{equation}
    |\Vec{u}_{p}(z)| = |\cos \polangle \sin \camangle - e^{i \Delta \beta z} \sin \polangle \cos \camangle|.
\end{equation}
Thereby, the predicted intensity captured by the camera is 
\begin{equation}
    I_1 (\polangle, z) \propto |{\Vec{u}_{p}(z)}|^2= 1 - \text{Re}\{e^{i 2 \polangle} W\},
\end{equation}
where we define $W= \cos 2 \camangle - i \sin 2 \camangle \cos \Delta \beta z$.
We observe that at every position $z$ along the fiber and for a fixed camera position $\camangle$, the intensity is a sinosoidal function of the polarization angle. 
In the experiment, we tune this angle by a half-waveplate in order to extract the interference visibility
\begin{equation}\label{eq:visibility_W}
    V_W = \frac{I_\text{max} - I_\text{min}}{I_\text{max} + I_\text{min}} = |W|.
\end{equation}

\bibliography{main}

\end{document}